\documentclass{pasj00}

\usepackage{color}

\Received{yyyy/mm/dd}
\Accepted{yyyy/mm/dd}
\Published{$\langle$publication date$\rangle$}
\SetRunningHead{S.\ Takeuchi, K.\ Ohsuga, and S. Mineshige}{
A Novel Jet Model: Magnetically Collimated, Radiation-Pressure Driven Jet
}

\usepackage{times,bm}

\begin{document}

\title{A Novel Jet Model: Magnetically Collimated, Radiation-Pressure Driven Jet}
\author{Shun \textsc{Takeuchi},\altaffilmark{1, 2}
       Ken \textsc{Ohsuga},\altaffilmark{3}
        and
        Shin \textsc{Mineshige}\altaffilmark{1}
        }
\altaffiltext{1}{Department of Astronomy, Graduate School of Science, Kyoto University, Sakyo-ku, Kyoto 606-8502}
\altaffiltext{2}{Kwasan Observatory, Kyoto University, Yamashina-Ku, Kyoto 607-8471}
\altaffiltext{3}{National Astronomical Observatory of Japan, Osawa, Mitaka, Tokyo 181-8588}
\email{E-mail : shun@kusastro.kyoto-u.ac.jp}

\maketitle

\begin{abstract}
Relativistic jets from compact objects are ubiquitous phenomena in the Unvierse,
but their driving mechanism has been an enigmatic issue over many decades.
Two basic models have been extensively discussed: 
magnetohydrodynamic (MHD) jets and
radiation-hydrodynamic (RHD) jets.
Currently, the former is more widely accepted, since magnetic field
is expected to provide both the acceleration and collimation mechanisms, whereas
radiation field cannot collimate outflow.
Here, we propose a new type of jets, 
radiation-magnetohydrodynamic (RMHD) jets, based on our global
RMHD simulation of luminous accretion flow onto a black hole
shining above the Eddington luminosity.
The RMHD jet can be accelerated up to the relativistic speed 
by the radiation-pressure force and is collimated by the Lorentz force of a magnetic tower,
inflated magnetic structure made by toroidal magnetic field lines
accumulated around the black hole,
though radiation energy greatly dominates over magnetic energy.
This magnetic tower is collimated by a geometrically thick
accretion flow supported by radiation-pressure force.
This type of jet may explain relativistic jets from Galactic microquasars,
appearing at high luminosities.
\end{abstract}

\section{Introduction}
Astrophysical jets, highly collimated, high velocity plasma outflows,
are found in many astrophysical objects, including young stellar objects (YSOs),
microquasars, active galactic nuclei (AGNs), and gamma-ray bursts (GRBs).
Importantly, the jets (except for YSO jets) are, at least, mildly relativistic
and sometimes highly relativistic and are
collimated with an small opening angle which is only a few degrees.
It is widely accepted from the observations
that these relativistic jets 
are associated with accretion flows onto a central compact object, such as black-holes
(see review by \cite{MirRod99}).
Despite a long research history since the 1970's, 
how to collimate and accelerate astrophysical jets
still remains an open question
(e.g., \cite{Ree84};  \cite{Mei03}).
    
There are two models proposed and considered
for the driving mechanism of relativistic jets:
magnetohydrodynamic (MHD) jets 
which are driven by the magnetic process
(\cite{Lov76}; \cite{BlaPay82}; \cite{UchShi85}; \cite{HawBal02}) 
and radiation-hydrodynamic (RHD) jets 
which are driven by the strong radiation-pressure force
(\cite{BisBli77}; \cite{Egg+88}; \cite{Sik+96}; \cite{Oku+09}).
The former is very promising, 
since not only the acceleration but also the collimation
can be accounted for, 
whereas the latter cannot generally explain the collimation.
However, we should remark that most of the previous MHD disk-jet simulations assumed that
the underlying accretion flows are radiatively inefficient;
i.e., low-luminosity accretion flow
(\cite{Ich77}; \cite{NarYi94}).
By contrast, radiation field dominates the dynamics
of luminous (radiation-dominated) accretion flows,
from which highly relativistic jets are occasionally observed, especially
in the cases of Galactic microquasars.

Here, we examined the complex interactions between matter, radiation, and magnetic fields 
in the high-luminosity accretion flow and jet,
based on our global radiation-magnetohydrodynamic (RMHD) simulations.
Our goal is to give good physical explanation
why and how relativistic jets can be produced in luminous accretion flows
and what kinds of roles magnetic field plays in the situations
that radiation field energy dominates over matter and magnetic energy.
 
\section{Overview of the RMHD Simulations}
We analyzed the RMHD simulation data of \citet{Ohs+09}. 
Here, we outline the numerical method.
We extended the resistive MHD simulations performed by \citet{Kat+04},
and solved the equations including the radiation transfer terms
numerically in a 2.5-dimensional domain.
We used cylindrical coordinates $(r,\theta, z)$,
where $r$, $\theta$, and $z$ are the radial distance, the azimuthal angle, 
and the vertical distance, respectively.
The basic equations are the continuity equation,
\begin{equation}
 \frac{\partial \rho}{\partial t}
 + {\bm \nabla} \cdot (\rho {\bm v}) = 0,
 \label{mass_con}
\end{equation}
the equations of motion,
\begin{eqnarray}
 \frac{\partial (\rho {\bm v})}{\partial t}  +&& {\bm \nabla} \cdot
 \left( \rho {\bm v}{\bm v} - \frac{{\bm B}{\bm B}}{4 \pi} \right) \nonumber \\
  &&= - {\bm \nabla} \left( p_{\rm gas}+\frac{|{\bm B}|^2}{8\pi} \right) 
 +\frac {\chi}{c} {\bm F}_0 
 -\rho{\bm \nabla}\psi_{\rm PN},
 \label{mom}
\end{eqnarray}
the energy equation of the gas,
\begin{eqnarray}
 \frac{\partial e }{\partial t}
  +&& {\bm \nabla}\cdot(e {\bm v}) \nonumber \\ 
   &&= -p_{\rm gas}{\bm \nabla}\cdot{\bm v}
  +\frac{4\pi}{c^2}\eta_{\rm a} J^2 -4\pi \kappa B_{\rm bb} + c\kappa  E_0,
  \label{gase}
\end{eqnarray}
the energy equation of the radiation,
\begin{eqnarray}
 \frac{\partial E_0}{\partial t}  +&& {\bm \nabla}\cdot(E_0 {\bm v}) \nonumber \\
  &&= -{\bm \nabla} \cdot {\bm F_0} - {\bm \nabla} {\bm v}:{{\bf P}_0}
 + 4\pi \kappa B_{\rm bb} - c\kappa E_0,
 \label{rade}
\end{eqnarray}
and the induction equation,
\begin{equation}
 \frac{\partial B}{\partial t}
 ={\bm \nabla} \times
 \left({\bm v}\times {\bm B}-\frac{4\pi}{c}\eta_{\rm a}{\bm J} \right).
 \label{ind}
\end{equation}
Here, $\rho$ is the gas mass density,
$\bm{v}=(v_r, v_\varphi, v_z)$ is the flow velocity,
$c$ is the speed of light,
$e$ is the internal energy density of the gas,
$p_{\rm gas}$ is the gas-pressure,
$\bm{B}=(B_r, B_\varphi, B_z)$ is the magnetic field,
$\bm{J}(=c{\bm \nabla}\times {\bm B}/4\pi)$
is the electric current,
$B_{\rm bb}$ is the blackbody intensity,
$E_0$ is the radiation energy density,
${\bm F}_0$ is the radiative flux,
${ {\bf P}}_0$ is the radiation-pressure tensor,
$\kappa$ is the absorption opacity,
and $\chi$ is the total opacity,
respectively.
We adopted anomalous resistivity, $\eta_{\rm a}$ (\cite{YokShi94}),
the flux-limited diffusion (FLD) approximation 
to evaluate $\bm F_0$ and ${\bf P}_0$ (\cite{LevPom81}),
and the pseudo-Newtonian potential, $\psi_{\rm PN}$, 
to incorporate the general relativistic effects,
given by $\psi_{\rm PN} = -GM_{\rm BH}/(\sqrt{r^2+z^2}-r_{\rm s})$ (\cite{PacWii80}).
Here, $r_{\rm s}$ ($=2GM_{\rm BH}/c^2$) represents the Schwarzschild radius,
where $G$ is the gravitational constant
and $M_{\rm BH}$ is the mass of the black hole.
As the equation of state, we use the relation, 
$p_{\rm gas} = (\gamma -1)e = \rho k_{\rm B} T_{\rm gas}/\mu m_{\rm p}$, 
where $\gamma$ is the specific heat ratio,
$k_{\rm B}$ is the Boltzmann constant,
$T_{\rm gas}$ is the temperature of the gas,
$\mu$ themean molecular weight,
and $m_{\rm p}$ is the proton mass, respectively.

The computational domain is $3 \,r_{\rm s} \leq r \leq 103 \,r_{\rm s}$ and 
$0 \leq z \leq 91.6 \,r_{\rm s}$.
The grid spacing of the radial distance, $\Delta r$, and the vertical distance, $\Delta z$,
is uniform ($\Delta r = \Delta z = 0.2 \, r_{\rm s}$).
The reflection symmetry relative to the equatorial plane (with $z=0$) is assumed.
Free boundary conditions for matter and a magnetic field are adopted; 
i.e., the matter can freely go out, but cannot to come in,
and the magnetic field cannot change across the boundaries.
Radiation is set to go out of the boundaries with the radiative flux of $c E_0$,
except at the innermost boundary at $r = 3 r_{\rm s}$ and $z > 3 r_{\rm s}$,
through which no radiation can go out.
We also set a non-rotating stellar-mass black hole ($M_{\rm BH} = 10 M_{\odot}$) at the origin, 
and initially put a rotating torus at radii around $40 \, r_{\rm s}$
threaded with closed poloidal magnetic field (plasma-$\beta \equiv p_{\rm gas}/p_{\rm mag}= 100$),
and a non-rotating isothermal corona.
Here, $p_{\rm mag}$ ($=B^2/8\pi$) is magnetic pressure force.

\section{RMHD Jet Emerging from a Luminous Accretion Flow}
\begin{figure*}
 \begin{center}
 \FigureFile(170mm,110mm){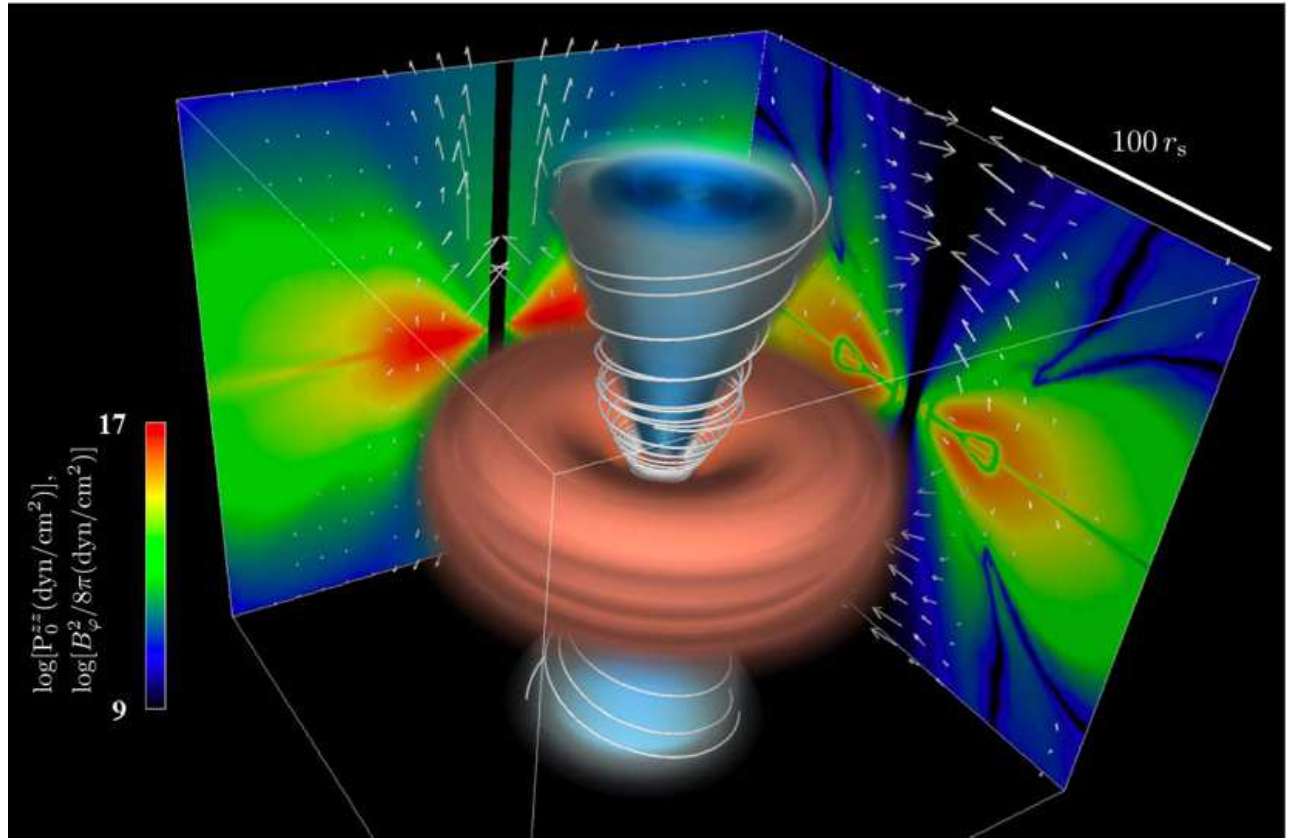}
 \end{center}
\caption{
Bird's-eye view of the luminous accretion flow and the associated RMHD jet.
The accretion flow (the gas mass density, brown)
and the RMHD jet in which velocities exceed the escape velocity
(the velocity, white, blue) are plotted.
The high-speed jet ($\sim 0.6 c - 0.7 c$) is represented by blue.
White lines indicate the magnetic field lines.
The $zz$-component of the radiation-pressure tensor (color), ${\rm P}_0^{zz}$,
overlaid with the radiation-pressure force vectors (arrows)
on the meridional plane is projected on the left wall surface,
while the magnetic pressure from the azimuthal component of the magnetic field (color), $B_\varphi^2/8\pi$,
overlaid with the Lorentz force vectors (arrows)
on the meridional plane is projected on the right wall surface.
The color bar corresponds to the logarithmic value of 
${\rm P}_0^{zz}$
and
$B_\varphi^2/8\pi$.
The arrows on each wall are displayed only in the region in which their values are larger than $10^{10} \, \rm dyn/g$.
Each quantity is time-averaged over 1 s,
which corresponds to the accretion timescale at several tens
of the Schwarzschild radius ($r_{\rm s}$).
}
\label{fig:jet1}
\end{figure*}

In Figure \ref{fig:jet1}, we show a bird's-eye view of
luminous accretion flow (at the center, with the brown color), outflow
(with the white and the blue color), and the associated magnetic field lines (with the white lines).
In the jet region only the parts with high velocities ($\sim 0.6 c - 0.7 c$) are
illustrated by the blue color.
The emergence of a magnetic tower formed by the accumulation of the toroidal magnetic field
which emerged from the underlying accretion flow is evident.
This is quite reminiscent of non-radiative MHD jets
(\cite{Lyn96}; \cite{Kat+04}; \cite{Nak+06}),
in spite of large radiation energy density which by far exceeds the magnetic energy
in the present case (discussed later).
From the simulation data, we obtain the following quantities;
the mass-outflow rate is $\dot{M}_{\rm out} \sim 10L_{\rm E}/c^2$;
the photon luminosity is $L_{\rm ph} \sim L_{\rm E}$;
the kinetic luminosity is $L_{\rm kin} \sim 0.1 L_{\rm E}$;
the Poynting luminosity is $L_{\rm poy} \sim 0.01 L_{\rm E}$.
Note that the Eddington luminosity ($L_{\rm E}$) can be exceeded in the case of
disk accretion onto black holes because of significant anisotropy
of the radiation field and photon trapping effects (\cite{OhsMin07}).
The mass-accretion rate to the central black hole is $\dot{M}_{\rm acc} \sim 100 \, L_{\rm E}/c^2$,
that exceeds the critical rate giving rise to the Eddington luminosity,
on the order of $10 L_{\rm E}/c^2$,
i.e., the simulated flow is supercritical
(\cite{ShaSun73}; \cite{Abr+88}; see Chap. 10 of \cite{Kat+08} for a review).
Here, the mass-accretion rate and the mass-outflow rate are
calculated by summing up the mass passing through the inner boundary
and the upper boundary per unit time with higher velocities than the escape velocity.
Photon luminosity, kinetic luminosity, and Poynting luminosity are
calculated based on the radiative flux, the kinetic energy, and magnetic energy at the upper boundary.

\begin{figure*}
 \begin{center}
 \FigureFile(170mm,110mm){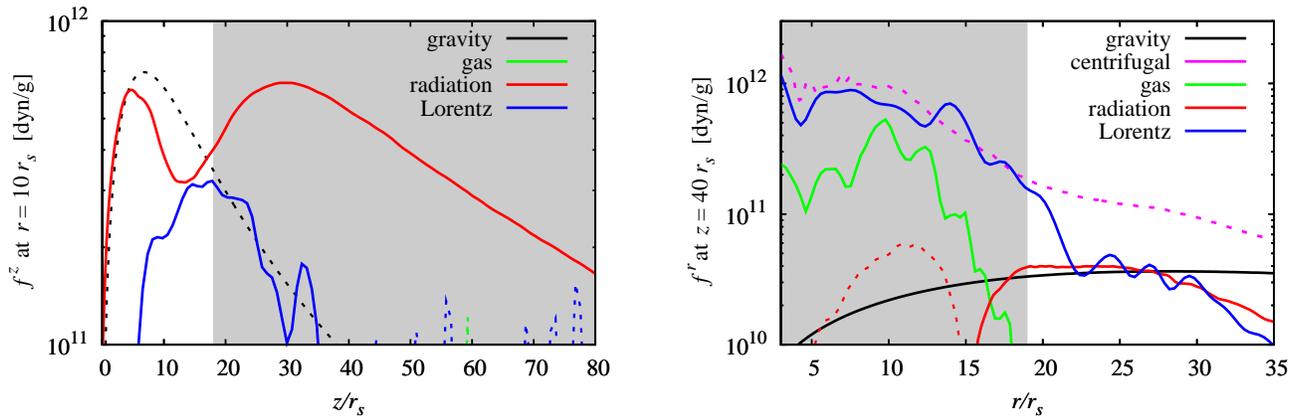}
 \end{center}
\caption{
Acceleration and collimation mechanism of the RMHD jet.
Left panel: The vertical profiles of
the gravitational force (black), 
the gas-pressure force (green), 
the radiation-pressure force (red), 
and the Lorentz force (blue)
at $r=10$ $r_{\rm s}$.
The solid lines and the dashed lines indicate the outward force and the inward force, respectively.
The radiation-pressure force is responsible for the jet acceleration.
Right panel: The radial profiles of
the gravitational force (black), 
the centrifugal force (magenta),
the gas-pressure force (green), 
the radiation-pressure force (red), 
and the Lorentz force (blue), 
at $z=40$ $r_{\rm s}$.
The solid lines and the dashed lines indicate the inward force and the outward force.
We understand that the Lorentz force is responsible for the collimation of the jet.  
In both panels each quantity is time-averaged over $1 \, \rm s$, and the
shadowed areas indicate the jet region in which the outflow velocity exceeds the escape velocity.
}
\label{fig:force}
\end{figure*}

\subsection{Acceleration Mechanism of RMHD Jet}
To see what force is responsible for the acceleration of the jet,
we evaluate the time-averaged strengths of the vertical components of the gravitational force, the gas-pressure force,
the radiation-pressure force, and the Lorentz force at a radius, $r=10 \, r_{\rm s}$,
and display them in the left panel of Figure \ref{fig:force} as functions of the vertical coordinate, $z$.
This figure unambiguously shows that the RMHD jet is accelerated by the radiation-pressure force
(indicated by the red line),
and that the Lorentz force (indicated by the blue line) does not contribute much to the acceleration.
This is a natural consequence, since the radiation energy density
greatly exceeds the magnetic energy density, typically, by more than one order of magnitude.
To understand the direction of the radiation-pressure force,
we display in the left wall surface of Figure \ref{fig:jet1}
the color contours of the $zz$-component of the radiation-pressure tensor,
${\rm P}_0^{zz}$,
overlaid with the radiation-pressure force vectors (arrows),
$\chi {\bm F_0} /\rho c$,
on the meridional plane.
We can clearly see that the vertical, outward component of the radiation-pressure force dominants in the jet region (around the $z$-axis),
since the radiation energy density rapidly varies in the vertical direction,
but not so much in the horizontal direction; that is, the significant anisotropy of the radiation field
(\cite{OhsMin07}).

\subsection{Collimation Mechanism of RMHD Jet}
Next, we examine the collimation mechanism of the RMHD jet
in the same manner as we did for the acceleration mechanism.
We calculate the time-averaged strengths of the radial components of
the gravitational force, gas-pressure force, centrifugal force, radiation-pressure force,
and the Lorentz force at $z=40$ $r_{\rm s}$,
and display them in the right panel of
Figure \ref{fig:force} as functions of $r$.
In contrast with the left panel,
the Lorentz force (indicated by blue solid line) dominates;
i.e., the magnetic force works as the a collimation force in the RMHD jet
against the outward (centrifugal) force (indicated by the magenta dashed line).
We note that magnetic-pressure force and the tension force are comparable.
To understand the direction of the Lorentz force in each region,
we display in the right wall surface of Figure \ref{fig:jet1} the color contours of
the magnetic pressure from the azimuthal component of the magnetic field,
$B_\varphi^2/8\pi$,
overlaid with the Lorentz force vectors (arrows),
${\bm J \times \bm B} /\rho c$,
of the RMHD flow on the meridional plane.

As we already mentioned,
the radial, inward component of the Lorentz force, 
i.e., pinch effect of the magnetic field, dominates over other forces,
 including the radiation-pressure force, in the jet region.
This seems, at first glance, odd, since the magnetic energy density is
much less than the radiation energy density; say, by a factor of $\sim 20$
at $(r,z)=(18 \; r_{\rm s}, 40 \; r_{\rm s}$).
Nevertheless, magnetic force can dominate over the radiation force,
since the radiation-pressure force is largely 
attenuated in the horizontal direction (in the disk plane)
because of highly anisotropic distribution of radiation energy.
In the high luminosity flow with large matter column density,
the accumulation of material around the black hole is responsible for generating rather flat distribution
of the radiation energy density in the horizontal direction.
As a result, the horizontal component of the radiation-pressure force,
which is proportional to the horizontal gradient of the radiation energy density,
is largely reduced. 
By contrast, magnetic energy density shows high concentration ("magnetic tower")
in the cylindrical region surrounding the black hole, asserting
large inward magnetic force at the inner surface of the magnetic tower.
The magnitude of the Lorentz force 
can be by $\sim 5$ times as large as that of the radiation force there.

Note that a magnetic field is by nature intrinsically expansive. 
Thus, there must be something to prevent the field from expanding 
sideways, thereby decollimating the whole configuration
(\cite{UzdMac06}; \cite{Spr10}).
In our simulations, 
it is the geometrically thick accretion flow supported by strong
radiation pressure force that prevents the filed from expansion. 
We can prove this by looking at Figure 2 (right) that the central
region (the magnetic tower) is confined by the combination of the 
radiation pressure force (the red curve) and the gravitational force 
(the black line). Cautions should be taken for the bird's eye view 
in Figure 1; it only traces very high density regions ("core" of the 
accretion flow) and substantial amount of material exists around 
the magnetic tower, as is illustrated in the left wall in Figure 1.

We started the simulations from the particular initial conditions,
and we admit that our results may depend on them.
We need further simulation studies starting from different initial 
magnetic field configurations(\cite{Igu+03}; \cite{Bec+09}). 
Cases of continuous addition of 
a magnetic field in a flow should also be necessary to examined
(\cite{MckNar07}).
We are planning such simulations in future work.

\section{Discussion}
Recently, \citet{Fen+04} compiled 
the X-ray observational data of Galactic microquasars
in the luminosity-hardness ratio diagram
and reported a evolutionary paths in the diagram.
Importantly, they found
a common condition for the occurrence of relativistic jets 
in microquasars; that is, 
relativistic jets only appear when the source exhibits a hard-to-soft transition 
at high luminosities.
The typical photon luminosities ($L_{\rm ph}$), at which relativistic jets are observed, are
close to (or, at least, a few ten percents of) $L_{\rm E}$.
Our results are consistent with such observational facts.

The present RMHD simulation, however, does not produce highly relativistic jet 
with the bulk Lorentz factor of a few.
This is partly because our simulations are restricted to the Newtonian dynamics.
In future relativistic RMHD simulations will able to show really
relativistic jets by extracting energy from the black hole spin energy.
Recently made general relativistic MHD simulations can indeed achieve high Lorentz factor of jets
(\cite{BlaZna77}; \cite{Koi+02}; \cite{Mck06}),
though radiation processes are not considered in their simulations.
In this sense, our simulations are more relevant to
mildly relativistic, persistent jets from a peculiar source, SS433.  In fact supercritical accretion is indicated in this source
 by number of observations
(\cite{Mar+79}; \cite{Kot+96}; \cite{Fab04}).

To conclude,
complex interaction between matter, radiation, and magnetic field would be an important factor for the relativistic jet formation at high luminosities.  Although radiation energy dominates, magnetic fields manifest themselves
in the form of magnetic tower structure, which provides a collimation mechanism,
which radiation field cannot give.
Such situations are expected to always occur at above 10\% of the Eddington luminosity,
above which the radiation energy density exceeds, or is at least comparable to, internal gas energy density, and thus exceeds the magnetic energy density.
Our global RMHD simulations demonstrate the universal role of 
magnetic field, as well as the importance of radiation-hydrodynamics in physics of 
black hole accretion disks and flows.

\bigskip
We would like to thank K. Shibata and Y. Kato for useful comments and discussions. 
This work is supported in part by the Grant-in-Aid of MEXT (19340044, SM), and by the Grant-in-Aid for the global COE programs on gThe Next Generation of Physics, Spun from Diversity and Emergenceh from MEXT (SM),
Ministry of Education, Culture, Sports, Science, and Technology (MEXT)
Young Scientist (B) 20740115 (KO).
Numerical computations were in part carried out on Cray XT4 at Center for Computational Astrophysics, CfCA, of National Astronomical Observatory of Japan.

\end{document}